\documentclass[epj,final]{svjour}

\usepackage{graphics}
\usepackage{amsmath}
\usepackage{amssymb}
\usepackage[usenames,dvipsnames]{color}
\newcommand{\ue}{\mathrm{e}}
\newcommand{\ud}{\mathrm{d}}
\newcommand{\pstar}{\mathcal{P}^*}
\newcommand{\Real}{\mathrm{Re}}
\newcommand{\im}{\mathrm{i}}
\newcommand{\h}{\mathrm{H}}
\newcommand{\ui}{\mathrm{i}}

\begin{document}
\title{Accounting for risk of non linear portfolios}
\subtitle{A novel Fourier approach}
\author{Giacomo Bormetti\inst{1,2} \and Valentina Cazzola\inst{1,3,2}% etc
\and Danilo Delpini\inst{3,2} \and Giacomo Livan\inst{3,2}
\mail{Giacomo.Bormetti@pv.infn.it}%
}                     % Do not remove
\institute{
Centro Studi Rischio e Sicurezza, Istituto Universitario di Studi Superiori,
V.le Lungo Ticino Sforza 56, Pavia, 27100, Italy
\and
Istituto Nazionale di Fisica Nucleare - Sezione di Pavia,
via Bassi 6, Pavia, 27100, Italy
\and
Dipartimento di Fisica Nucleare e Teorica, Universit\`a degli Studi di Pavia,
via Bassi 6, Pavia, 27100, Italy
}
\date{Received: date / Revised version: date}
\abstract{
The presence of non linear instruments is responsible for
the emergence of non Gaussian features in the price changes distribution of
realistic portfolios, even for Normally distributed risk factors.
This is especially true for the benchmark Delta Gamma Normal model, which
in general exhibits exponentially damped power law tails.
We show how the knowledge of the model
characteristic function leads to Fourier representations for two standard risk
measures, the Value at Risk and the Expected Shortfall, and for their sensitivities with
respect to the model parameters.
We detail the numerical implementation of our formulae and
we emphasizes the reliability and efficiency of our results in comparison with Monte Carlo simulation.
} %end of abstract
\maketitle
\section{Introduction}
\label{intro}
In the last two decades, the Econophysics community has deeply investigated the
empirical features of historical time series emerging in different financial
contexts,
ranging from high to low frequency data from the stock exchange to the markets
in which bonds, FX rates, commodities, energy, futures, options and many other
instruments are traded. 
Numerous empirical evidences have emerged, sometimes confirming results from the
financial-econometrics community, or finding new stylized facts and opening new fields of research
\cite{Mantegna_Stanley:2000,Bouchaud_Potters:2003,Voit:2001,McCauley:2004}.  
A well known example is given by the observation that the intra-day or daily
stochastic dynamics of stock prices significantly deviates from the behaviour
predicted by the geometric Brownian motion.
This finding dates back to the work of Mandelbrot
\cite{Mandelbrot:1963}, whose attention was mainly focused in recognizing
realizations of stable processes, and to the analysis of Fama~\cite{Fama:1965}
concerning the long tailed nature of the Dow Jones Industrial Average single components.
The statistics of returns has been recently rediscussed in different flavours and
its modelling has considerably grown. Very heterogeneous models, able to reproduce the
degree of asymmetry and the excess of kurtosis of the measured distributions, have been
proposed. 
A non exhaustive list includes approaches developing from specific distributional
assumptions, as it is the case of the 
L\'evy flights~\cite{Mantegna_Stanley:2000,Mantegna:1991,Mantegna_StanleyPRL:1994},
the Generalized Student-$t$ or 
Tsallis distributions~\cite{Bouchaud_Potters:2003,BorlandPRL:2002,Bormetti_etal:2007}
and the exponential one~\cite{McCauley_Gunaratne:2003}, 
and it can be extended to more sophisticated models capturing the stochastic nature
of the volatility, see~\cite{Fouque_etal:2000,Dragulescu_Yakovenko:2002,Perello_etal:2004,Bormetti_etal:2008} and references therein,
the power law scaling of its autocorrelation function and the leverage
effect~\cite{Bouchaud_Potters:2003,Perello_Masoliver:2003} 
and the multi-fractal properties of historical time
series~\cite{Muzy_etal:2000,Pochart_Bouchaud:2002}. The level of complexity grows
increasing the dimensionality of the problem. 
The leptokurtic nature of returns distributions has been shown also for financial
indexes, which are linear combinations of
plain stocks, such as the Standard \& Poor's 500~\cite{Mantegna_Stanley:2000,Mantegna:1995}
or the NIKKEI~\cite{Gopikrishnan:1999}.
From the point of view of the Central Limit Theorem, this result was not expected:
because of the great number of components, the convergence towards the Gaussian
regime would be ensured also for heavy tailed stock returns with finite variance.
Moving back from the index to the components has proved that
the single assets themselves share the same power law scaling, with tail index close
to 3~\cite{Gopikrishnan2:1999}. Indeed, this fact would be responsible for an ultra
slow convergence to Normality for independent stocks; however the existence of strong
correlations among assets is a well known evidence which violates the assumptions of
the Central Limit Theorem.

In this paper we investigate a completely different mechanism at work,
even when the single factors governing the aggregate behaviour are assumed to
be Gaussian. We move from the analysis of financial indexes to the case
of portfolios where non linear instruments, such as option
contracts, induce deviations from Normality, which are no more a consequence
of the microscopic dynamics. In the following sections we will review the standard Delta
Gamma Normal (DGN) approach to the problem of risk management for non linear portfolios
and we will discuss a novel analytical methodology and its efficient numerical
implementation.
In particular, our aim is to evaluate two standard risk measures, the Value-at-Risk
(VaR) and the Expected Shortfall (ES) in the framework of the DGN approximation.
For an introduction we refer the reader to the standard financial literature
~\cite{Jorion:2001,Mina:2001,Acerbi:2002}. As far as the single asset case is concerned,
many different approaches can be adopted. A widely exploited technique in risk analysis, requiring no
assumptions on the distribution of the data consists in using past returns realizations
to compute risk exposure levels, which are associated to the empirical quantiles of
the distribution. Despite its simplicity and generality, such approach suffers from
the typical drawbacks related to finite size effects. On the other hand,
if an assumption is made on the data distribution, various parametric approaches
are available, both in a frequentist and Bayesian framework, often leading to analytical
or semi-analytical expressions for VaR and ES~\cite{Mina:2001,Kondor:2001}.
Beyond the standard results for Normally distributed asset returns, in Econophysics
risk measures have been evaluated with heavy tailed probability density functions (PDF),
see for example~\cite{Bouchaud_Potters:2003} and approaches based on
generalized Student-$t$~\cite{Bormetti_etal:2007}, Tsallis~\cite{Mantegna:2004}
and Truncated L\'evy random variables~\cite{Bormetti:2010}. Equivalent estimates to
those for i.i.d. Student-$t$ variables can also be recovered in a different approach
exploiting Bayesian Product Partition Models for Normal independent but not identically
distributed returns~\cite{Bormetti2:2010}.
Furthermore, advanced approaches based on the so-called downside risk and applied to
the specific case of hedge funds can be found in~\cite{Lhabitant:2002,Perello:2007}.
In this paper, we follow an alternative approach based on the generalized Fourier
representation of the PDF, already developed
in~\cite{Bormetti:2010} for the single asset case. Fourier techniques for
VaR evaluation have been introduced since the work of Rouvinez~\cite{Rouvinez:1997},
and recently Martin~\cite{Martin:2009} provided an extension of
this framework, obtaining expressions also for the ES and its
sensitivities. However, despite being in semi-closed form, they are not
straightforwardly suited to numerical evaluation but
require a saddle point approximation. On the contrary our analytical formulae can be
easily and efficiently evaluated by means of standard trapezoidal integration or Fast
Fourier algorithms.

This article is organized as follows. In section~\ref{sec:Risk_measures}
we review the portfolio model we consider, outlining the available analytical
information (its characteristic function and the corresponding asymptotic tail
behaviour of the PDF). In that section we derive the semi-analytical expressions
for the VaR, ES, and their sensitivities with respect to the model parameters,
which represent the original contribution of this work. In section~\ref{sec:Numerical}
we compare the numerical results obtained through our expressions with those
from the Monte Carlo simulation of synthetic portfolios.
Eventually, the relevant conclusions are drawn in section~\ref{sec:conclusions}.

\section{A first step beyond the Normal behaviour}
\label{sec:Risk_measures}
A realistic financial situation usually involves portfolios containing relevant
quantities of non linear (options) instruments. A description of these portfolios
in terms of a linear composition of risk factors is inadequate, especially when
assessing the market risk exposure. An improvement 
consists in taking a second order expansion of portfolio variations in the risk
factors. In particular, when these are assumed to be Normally distributed,
one obtains the DGN model. Since this model has become a benchmark
framework in the common practice of portfolio risk management, 
in the present work we do not discuss the limits of applicability of the DGN approximation,
which are analyzed in~\cite{Mina:2001,Mina:1999,Britten-Jones:1999}.

\subsection{The Delta Gamma Normal approach}
According to the DGN model, the portfolio price change $V$ over a
given time horizon $\Delta t$ is described by the quadratic form
\begin{equation}
  V = \theta + \Delta^{\top} X + \frac{1}{2} X^{\top} \Gamma X \ ,
  \label{DGN}
\end{equation}
where $X$ is the $N$-dimensional vector of the risk factors
which are responsible for the portfolio fluctuations. Here,
the variation is defined as $V=W-W_0$, where $W_0$ denotes the value of
the portfolio at the present time $t=t_0$, while $W$ is the corresponding
value at $t=t_0+\Delta t$.
The DGN model assumes $X$ to be drawn from a multivariate Gaussian distribution $\mathcal{N}(0,\Sigma)$ with
zero mean and covariance matrix $\Sigma$. In equation~\eqref{DGN} $\theta \in \mathbb{R}$
and $\Delta \in \mathbb{R}^N$ are constants, and so is the real symmetric $N \times N$ matrix
$\Gamma$ which accounts for possible non linearities.

By solving the generalized eigenvalue problem
\begin{equation}
  CC^{\top} = \Sigma \quad\mathrm{and}\quad C^{\top} \Gamma C = \Lambda\ ,
  \label{eigprob}
\end{equation}
with $\Lambda = \mathrm{diag} (\lambda_1, \ldots, \lambda_N)$, equation~\eqref{DGN} can be conveniently rewritten as
\begin{equation}
  V = \theta + \sum_{i = 1}^N \left ( \delta_i Y_i + \frac{\lambda_i}{2} Y_i^2 \right )\ ,
  \label{indipDGN}
\end{equation}
where $\delta = C^{\top} \Delta$ and $X = CY$. The $Y_i$'s now represent independent standard
Gaussian variables, and the correlation structure between the actual risk factors contained in
$X$ is now spread across the new $\delta$ and $\lambda$ parameters through relations~(\ref{eigprob}).

Interestingly, the moment generating function of the random variables appearing in equation
(\ref{indipDGN}) has been computed in~\cite{Jaschke:2002} as a Gaussian integral, yielding 
\begin{equation}
  \mathbb{E} [\ue^{\omega (\delta_i Y_i + (\lambda_i Y_i^2) / 2)}] = \frac{1}{\sqrt{1 - \lambda_i \omega}}
  \exp \left \{ -\frac{1}{2} \frac{\delta_i^2 \omega^2}{1 - \lambda_i \omega} \right \}\ ,
  \label{mgfunction}
\end{equation}
$\omega \in \mathbb{R}$, and since this is a holomorph function in the neighbourhood of $0$, the result can be extended
to the complex domain. So, the characteristic function $f$ of (\ref{indipDGN}) can be computed:
\begin{equation}
  f(\omega) = \mathbb{E} [\ue^{\ui \omega V}] = \ue^{\ui \theta \omega} \prod_{i=1}^N \frac{1}{\sqrt{1 - \ui \lambda_i \omega}}
  \exp \left \{ -\frac{1}{2} \frac{\delta_i^2 \omega^2}{1 - \ui \lambda_i \omega} \right \}\ .
  \label{charf}
\end{equation}
Risk assessment applications typically require a careful estimation of the tails. Even though not required by
the VaR and ES evaluation presented in this paper, we devote the last part of the present subsection to review the analytical results
for the tails of the PDF of the DGN model, which
have been fully characterized by Jaschke \emph{et al.} in~\cite{Jaschke:2004}.
Indeed, let us suppose that the $\lambda_i$
eigenvalues have been ordered in increasing order, and let $n \leq N$ be the number of distinct values. Let $i_k$ be
the highest index of the $k$-th distinct eigenvalue (so that $\lambda_{i_1} < \ldots < \lambda_{i_n}$)
and let $m_k$ be its multiplicity. Then we can define the following useful quantities
for $k=1,\cdots,n$
\begin{equation*}
  \bar{\delta}_k^2 = \sum_{j = i_{k-1} + 1}^{i_k} \delta_j^2 \quad\mathrm{and}\quad
  a_k^2 = \frac{\bar{\delta}_k^2}{\lambda_{i_k}^2}\ .
\end{equation*}
The left tail behaviour of the PDF $p(V)$ is determined by the sign of the lowest
eigenvalue $\lambda^*\doteq\lambda_{i_1}$.
When $\lambda^* < 0$, the left tail displays an exponentially damped power law decay,
whose rate is actually given by $\lambda^*$. More precisely, when
$V \rightarrow - \infty$, $p(V)$ has the following asymptotic tail behaviour
\begin{align}
  p(V) &= c(\delta,\lambda) \ |V|^{\bar{m}} \ \ue^{V /
  \lambda^* + a_1 \sqrt{| 2V / \lambda^* |} }  \nonumber \\
  &\times \big (1 + O(1 / \sqrt{|V|}) \big )\ ,
  \label{eq:Jaschke_m}
\end{align}
where
\begin{equation*}
  \bar{m} = \left \{ \begin{array}{ll}
    (m_1 - 3)/4 & \ \ \ \mathrm{if}  \ \ a_1 \ne 0\ \\
    m_1/2 -1 & \ \ \ \mathrm{if} \ \  a_1 = 0\ .
  \end{array} \right.
\end{equation*}
When $\lambda^* = 0$, the left tail is characterized by an asymptotically Gaussian scaling:
\begin{align}
  p(V) &= d(\delta,\lambda) \ |V-V_0|^{-\frac{1}{2} \sum_{k=2}^n m_k}
  \ue^{-(V-V_0)^2 / (2 \bar{\delta}_1^2)} \nonumber\\
  &\times \big (1 + O(1/|V-V_0|) \big )\ ,
  \label{eq:Jaschke_z}
\end{align}
where $V_0 = \theta - \sum_{k=2}^n \bar{\delta}_k^2 / (2 \lambda_{i_k})$.
It is worth mentioning that, both in this case and the previous one, the contribution coming from the power law factor
may not be negligible when estimating relevant quantiles; moreover, in this context deviations from
Gaussianity arise as a consequence of the inclusion of second order terms in the portfolio
variation~\eqref{DGN}, and not of the assumption of a fat tailed distribution for
the risk factors.

Lastly, when $\lambda^* > 0$ the PDF is zero for $V \leq V_{\mathrm{inf}}$,
with $V_{\mathrm{inf}} = \theta - \sum_{k=1}^n \bar{\delta}_k^2 / (2 \lambda_{i_k})$, and
it decays as a power law in the limit $V \rightarrow V_{\mathrm{inf}}^+$:
\begin{equation}
  p(V) = k(\delta,\lambda) \ (V-V_{\mathrm{inf}})^{N/2-1} \big ( 1 + O(V-V_{\mathrm{inf}}) \big )~.
  \label{eq:Jaschke_p}
\end{equation}
In equations~\eqref{eq:Jaschke_m},~\eqref{eq:Jaschke_z},~\eqref{eq:Jaschke_p},
$c(\delta,\lambda) = c(\delta_{i=1,\ldots,N},\lambda_{i=1,\ldots,N})$,
$d(\delta,\lambda)$ and $k(\delta,\lambda)$ are constants depending on the set
of parameters in use.

As far as risk estimation is concerned, the left tail decay is what we are
interested in; however, previous considerations apply to the right tail
in an antithetic way depending on the sign of the highest eigenvalue $\lambda_{i_n}$.
When $\lambda_{i_n}>0$ ($\lambda_{i_n}=0$) the right tail is exponential (Gaussian) in the
limit $V\to +\infty$; otherwise, the support of $p(V)$ is limited from the right
and for $V\to V_{\mathrm{sup}}^-$ it scales as a power law with exponent $N/2-1$ and 
$V_{\mathrm{sup}}=\theta-\sum_{k=1}^n \bar{\delta}_k^2 / (2 \lambda_{i_k})$.

The central moments of the distribution of $V$ can be also explicitly derived, and the first
four of them read~\cite{Jaschke:2002,Mina:1999,Britten-Jones:1999}:
\begin{align*}
  \mu_1 &= \theta + \frac{1}{2} \mathrm{tr}[\Gamma \Sigma] \\ \nonumber
  \mu_2 &= \Delta^{\top} \Sigma \Delta + \frac{1}{2} \mathrm{tr} [(\Gamma \Sigma)^2] \\ \nonumber
  \mu_3 &= 3\Delta^{\top} \Sigma \Gamma \Sigma \Delta + \mathrm{tr} [(\Gamma \Sigma)^3] \\ \nonumber
  \mu_4 &= 12 \Delta^{\top} \Sigma (\Gamma \Sigma)^2 \Delta + 3 \ \mathrm{tr} [(\Gamma \Sigma)^4] 
  + 3 \mu_2^2\ .
\end{align*}
From these relations explicit expressions for the skewness $\zeta = \mu_3 / \mu_2^{3/2}$ and the kurtosis
$\kappa = \mu_4 / \mu_2^2-3$ can be derived, and one can easily check that whenever $\Gamma = 0$
we have $\zeta = 0$ and $\kappa = 0$. This is coherent with the fact that equations (\ref{DGN})
and (\ref{indipDGN}) define a Gaussian portfolio model when their quadratic terms are set to zero,
i.e. when $\Gamma = 0$. From this point of view, the possible presence of asymmetries or non Gaussian tails in
the PDF of the DGN model stems from the non linear terms in (\ref{DGN}) and (\ref{indipDGN}).

\subsection{Formulae for risk estimation}
The risky nature of a portfolio can be accounted for via the well-known VaR
estimator, which gives the potential loss (negative variation) over the time horizon
$\Delta t$ that could be exceeded with probability equal to the significance level $\pstar$.
However the VaR is known to suffer from two main drawbacks, the lack of subadditivity and
of information about the average potential loss when the VaR threshold is exceeded.
Both of them are overcome by introducing an alternative and coherent measure~\cite{Acerbi:2002}
known as the ES.

To obtain semi-analytical expressions for the risk measures, the steps followed
in~\cite{Bormetti:2010} are substantially replied.
We fix a significance level $\pstar \in (0,1)$ and we define the VaR $\Delta^*$ as
\begin{align}
  \pstar &= \int_{-W_0}^{-\Delta^*} \ud V \ p(V) \nonumber\\
  &= \int_{-\infty}^{-\Delta^*} \ud V ~ p(V)
  -\int_{-\infty}^{-W_0} \ud V ~ p(V)\ .
  \label{VaR}
\end{align}
In this expression $-W_0$ represents the maximum possible loss over $\Delta t$. 
Compared to~\cite{Bormetti:2010}, this framework involves portfolio variations
instead of logarithmic returns and this leads to the finite lower bound of
integration in the first line of equation~(\ref{VaR}).
Since we are modelling the risk factors as Gaussian random variables,
$p(V)$ could have unbounded support; nevertheless, under the assumption that the DGN
$p(V)$ is a good approximation to the true (unknown) PDF, we expect $-W_0$
to be on the far left tail of $p(V)$ and the second term of
equation~\eqref{VaR} to be negligible~\footnote{The second
term in equation~\eqref{VaR} can be evaluated numerically given $W_0$. Should it be not
negligible, the expressions presented in this paper are extended in a straightforward
way by taking into account the surface terms from
integrations.}. For this reason, in the following
expressions, we always imply the limit $-W_0\to -\infty$.

We now represent $p(V)$ in terms of its generalized Fourier transform
\begin{equation}
  p(V) = \frac{1}{2\pi}\int_{-\infty + \ui \nu}^{+\infty + \ui \nu} \ud \phi ~
  f(\phi)~\ue^{- \ui \phi V} \ ,
  \label{GenFourT}
\end{equation}
where the axis of integration in the complex plane has to be chosen parallel to the real axis with imaginary part $\nu$ belonging to the strip
of regularity $(\nu_-,\nu_+)$ of $f$~\cite{Lewis:2001,Lipton:2001}.
The boundaries of the strip of regularity are determined by the singular points of $f$ closest to the origin.
If any, singularities must be purely imaginary, and, for the present case, a quick analysis shows that they are equal to 
$\{-\ui/\lambda_{i_k},\  k=1,\dots,n\}$. 
Using the expression~\eqref{GenFourT} and switching the
integration order, equation~(\ref{VaR}) becomes
\begin{equation*}
  \pstar = \frac{1}{2\pi} \int_{-\infty + \ui \nu}^{+\infty + \ui \nu} \ud \phi
  ~f(\phi) \left( \int_{-\infty}^{-\Delta^*} \ud V \ue^{-\ui \phi V} \right)
\end{equation*}
and the convergence of the innermost integral is guaranteed if we restrict $\nu\in (0,\nu^+)$.
With this choice, the previous expression readily reduces to
\begin{align}
  \pstar &= \frac{\ui}{2\pi} \int_{-\infty + \ui \nu}^{+\infty + \ui \nu} \ud \phi \
  \frac{f(\phi)}{\phi} \ \ue^{-\ui \phi \Delta^*} \nonumber  \\
  &= \frac{\ue^{-\nu \Delta^*}}{\pi} \Real \left [ \int_0^{+\infty} \ud \omega \
  \frac{f(\omega + \ui \nu)}{\nu - \ui \omega} \ \ue^{\ui \omega \Delta^*} \right ]\ ,
  \label{VaRFFT}
\end{align}
where $\phi=\omega+ \ui \nu$.
From the previous discussion, we can determine the value of $\nu^+$ depending on
that of $\lambda^*$:
\begin{itemize}
  \item when $\lambda^*<0$, then $\nu_+ = |1/\lambda^*|$;
  \item when $\lambda^*\ge 0$, then $\nu_+ = +\infty$.
\end{itemize} 
To evaluate the ES the considerations already exposed for VaR evaluation still apply.
Starting from its definition and using again equation~\eqref{GenFourT}, we can write
\begin{align*}
  E^*(\pstar) &= - \frac{1}{\pstar} \int_{-\infty}^{-\Delta^*} \ud V \ V \ p(V)\\ \nonumber
  &= - \frac{1}{2 \pi \pstar} \int_{-\infty+\ui \nu}^{+\infty+\ui \nu} \ud \phi
  \ f(\phi) \left [ \frac{1}{\phi^2} - \frac{\ui \Delta^*}{\phi} \right ]
  \ \ue^{\ui \phi \Delta^*} \\ \nonumber 
  &= - \frac{\ue^{-\nu \Delta^*}}{\pi \pstar} \Real \left [ \int_0^{+\infty} \ud \omega
  \ \frac{f(\omega+\ui\nu)}{(\omega + \ui \nu)^2} \ \ue^{\ui \omega \Delta^*} \right ] \\ \nonumber
  & + \frac{\ue^{-\nu \Delta^*}}{\pi \pstar} \Delta^*
  \Real \left [ \int_0^{+\infty} \ud \omega \ \frac{f(\omega+\ui \nu)}{\nu - \ui\omega}
  \ \ue^{\ui\omega \Delta^*} \right]~,
\end{align*}
and, recalling the representation~\eqref{VaRFFT} of $\pstar$, the last expression can be
simplified in
\begin{equation}
  E^*(\pstar) = \Delta^* - \frac{\ue^{-\nu \Delta^*}}{\pi \pstar}
  \Real \left [ \int_0^{+\infty} \ud \omega \ \frac{f(\omega+\ui\nu)}{(\omega + \ui \nu)^2} \
  \ue^{\ui \omega \Delta^*} \right ] ~.
  \label{eq:ES}
\end{equation}

\subsection{Sensitivities}
We are not only interested in estimating VaR and ES, but also in their
sensitivity with respect to the parameters of the DGN model.
For a portfolio which is a linear composition of risk factors, $\Delta^*$ and $E^*$
are 1-homogeneous functions~\cite{Martin:2009,Major:2004}. 
Indeed, when $\Gamma=0$, we have $\Delta= \delta$ and, posing $\theta= 0$ for simplicity, the portfolio VaR
reduces to the well known expression
\begin{equation*}
  \Delta^* = -\sqrt{\delta^T \Sigma \delta}~
  \sqrt{2}~\mathrm{erf}^{-1} (2\pstar-1)~,
\end{equation*}
where $\mathrm{erf}^{-1}$ is the inverse of the error function. The first derivatives of VaR read
\begin{equation*}
  \frac{\partial\Delta^*}{\partial \delta_k} = -
  \frac{\left(\Sigma \delta\right)_k}
  {\sqrt{\delta^T \Sigma \delta}}~
  \sqrt{2}~\mathrm{erf}^{-1} (2\pstar-1)~,
\end{equation*}
and the homogeneity condition is readily verified
\begin{equation*}
  \Delta^* = \sum_{k=1}^{N} \delta_k \frac{\partial\Delta^*}{\partial \delta_k}~.
\end{equation*}
It can be shown that an analogous relation holds for the $E^*$.
The main consequence is that for linear portfolios,
the knowledge of the first derivatives allows, just by itself, for a
complete reconstruction of the risk measures. Thus, risk measures response
with respect to shocks in the $\delta_k$ weights defining the portfolio
composition can be fully characterized with no need of computing higher
order derivatives.

On the other hand, in this work we consider the effects due to the presence of non linear
instruments; in this scenario, a complete description of the market risk exposure would
require the knowledge of higher order derivatives. Nevertheless, the first order
ones still provide crucial information about the sensitivity of risk
measures when shocking the portfolio weights. For this reason, here we 
report the expressions of the first order derivatives only, reminding the reader
that higher order terms can be computed going back through the same 
passages.

Equation~(\ref{VaRFFT}) is now differentiated with respect to 
the generic parameter that, for the sake of tidiness, we address as $\beta$
\begin{equation*}
  \{ \beta \} = (\theta, \delta_i, \lambda_i) \ \ \ i = 1, \ldots, N\ .
\end{equation*}
Since we aim at evaluating the response of the risk measures
with respect to shocks to the portfolio parameters while keeping $\pstar$ fixed, we obtain 
\begin{align*}
  0 = \frac{\partial \pstar}{\partial \beta} = \frac{\ue^{-\nu \Delta^*}}{\pi}
  &\Real \left [ \int_0^{+\infty} \frac{\ud \omega}{\nu - \ui \omega} \right.
  \left ( \frac{\partial f(\omega + \ui\nu)}{\partial \beta} \right. \\ \nonumber
  &\left. \left. + (\ui \omega - \nu) f(\omega+\ui\nu)
  \frac{\partial \Delta^*}{\partial \beta} \right ) \right ].
\end{align*}
Exploiting the previous equality, and recalling equation~\eqref{GenFourT},
the VaR sensitivities read
\begin{align}
  \label{VaRderiv}
  \frac{\partial \Delta^*}{\partial \beta}
  &= \frac{\Real \left [ \int_0^{+\infty} \frac{\ud \omega}{\nu - \ui \omega}
  \frac{\partial f(\omega+\ui\nu)}{\partial \beta} \
  \ue^{\ui \omega \Delta^*} \right ]}{\Real \left [ \int_0^{+\infty} \ud \omega \
  f(\omega + \ui \nu) \ \ue^{\ui\omega \Delta^*} \right ]} \\ \nonumber
  &= \frac{\ue^{-\nu \Delta^*}}{\pi p(-\Delta^*)} \Real \left [ \int_0^{+\infty}
  \frac{\ud \omega}{\nu - \ui \omega} \frac{\partial f(\omega+\ui\nu)}{\partial \beta}
  \ \ue^{\ui \omega \Delta^*} \right ]~.
\end{align}
Finally, using again~\eqref{VaRFFT}, differentiation of equation~\eqref{eq:ES} gives
us the following Fourier representation of the ES sensitivities
\begin{align}
  \frac{\partial E^*(\pstar)}{\partial \beta} &= \frac{\partial \Delta^*}{\partial \beta} 
  \left \{ 1 -\frac{\ue^{-\nu \Delta^*}}{\pi \pstar} \right.\nonumber \\
  &\left. \times ~ \Real \left [ \int_0^{+\infty} \ud \omega \ \frac{f(\omega +\ui\nu)}{\nu-\ui\omega} \
  \ue^{\ui\omega \Delta^*} \right ] \right \} \nonumber \\
  &- \frac{\ue^{-\nu \Delta^*}}{\pi \pstar} \Real \left [ \int_0^{+\infty}
  \frac{\ud \omega}{(\omega +\ui\nu)^2} \frac{\partial f(\omega +\ui\nu)}{\partial \beta} \
  \ue^{\ui \omega \Delta^*} \right ] \nonumber \\
  =& - \frac{\ue^{-\nu \Delta^*}}{\pi \pstar} \Real \left [ \int_0^{+\infty}
  \frac{\ud \omega}{(\omega +\ui\nu)^2} \frac{\partial f(\omega +\ui\nu)}{\partial \beta}
  \ \ue^{\ui \omega \Delta^*} \right ]~,
  \label{ESderiv}
\end{align}
where the final equality holds because the expression in the curly brackets vanishes.
Finally we compute the derivatives of the characteristic function appearing in
(\ref{VaRderiv}) and (\ref{ESderiv}). Using equation~\eqref{charf} we obtain
\begin{align}
  \frac{\partial f (\phi)}{\partial \theta} &= \ui \phi f (\phi)\ , \nonumber \\
  \frac{\partial f (\phi)}{\partial \delta_i} &= - \frac{\delta_i \phi^2}{1-
  \ui \lambda_i \phi}~f (\phi)\ , \nonumber \\
  \frac{\partial f (\phi)}{\partial \lambda_i} &= \frac{\ui \phi}{2(1-\ui \lambda_i\phi)}
  \left [ 1-\frac{\delta_i^2 \phi^2}{1-\ui \lambda_i\phi} \right] f(\phi)\ .
  \label{derCF}
\end{align}
Incidentally, we notice that all the previous relations, valid for the DGN model, are linear in $f$ and
thus equations~\eqref{VaRderiv} and \eqref{ESderiv} share a similar structure 
with equations~\eqref{VaRFFT} and~\eqref{eq:ES}.

\section{Numerical results}
\label{sec:Numerical}
In this section we detail the numerical results of our Fourier approach, computing
$\Delta^*$, $E^*$ and the sensitivities for synthetic portfolios corresponding
to the possible cases depending on the value of the smallest eigenvalue $\lambda^*$.
We check the semi-analytical estimates with the ones obtained by Monte Carlo simulation of
the portfolio values. Below, a brief summary of the numerical setup for both the semi-analytical
approach, proposed here, and the standard historical simulation one is given.

We do not address the problem of estimating the
covariance matrix $\Sigma$ from real portfolio data. Nevertheless, it is
worth noticing that the usual parametric Maximum Likelihood estimators of dispersion
perform reasonably well in the limit of an infinite number of observations
for the risk factors. When the length of the time series $T$ is larger 
than the number of risk factors $N$, it is also possible to exploit filtering
procedures~\cite{Mantegna:1999,Tumminello:2007}, such
as hierarchical clustering, to extract the relevant information from
correlation matrices, retaining only the statistically significant correlations.
These techniques, besides reducing the dimensionality of the problem,
can give crucial information for decision processes such as portfolio
allocation. 
Under the thermodynamical limit of $T, N\rightarrow +\infty$, with $T/N$ fixed,  
the random matrix theory has also proved to be a useful
tool to better understand the statistical structure of empirical
correlation matrices~\cite{Laloux:1998,Plerou:1999}.
On the other hand, when the sample is small,
especially when $T$ is smaller than $N$, which is common for large portfolios,
the error affecting the Maximum Likelihood estimators is large. In this situations,
it is possible to resort to better performing estimators, such as, for instance,
the Ledoit and Wolf shrinkage estimator of dispersion~\cite{Ledoit:2003,Meucci:2005} which is a weighted average of
the sample covariance $\hat{\Sigma}$ and a target diagonal matrix $\hat{C}$
\begin{equation*}
  \hat{\Sigma}^S = (1-\alpha) \hat{\Sigma} + \alpha \hat{C}\ ,
\end{equation*}
where $\hat{C} = (1/m) \sum_{k = 1}^m \hat{\eta_k} \mathbb{I}$,
$\hat{\eta}_k$ are the sample estimations of the eigenvalues
of $\Sigma$ and $\alpha$ is the optimal shrinkage weight depending
on $\hat{\Sigma}$, $\hat{C}$ and $X_{t=1,\dots,T}$.

\subsection{Numerical setup}
\emph{Fourier inversion}. The complex integrals involved by our expressions
for risk measures and sensitivities are evaluated by means of trapezoidal
integration; this requires to split explicitly the real part of the integrand,
being an even function of $\omega$, in the two terms proportional to the sine and
cosine functions. For instance, evaluation of $\pstar$, equation~\eqref{VaRFFT},
reduces to compute the following
two real Fourier integrals
\begin{align*}
  \pstar = \frac{\ue^{-\nu \Delta^*}}{\pi}
  &\left\{ \int_0^{+\infty} \Real \left[ \frac{f(\omega+\im \nu)}{\nu - \im \omega} \right]
  \cos(\omega \Delta^*) \right.\\
  &- \left. \int_0^{+\infty} \mathrm{Im}\left[ \frac{f(\omega+\im \nu)}{\nu - \im \omega} \right]
  \sin(\omega \Delta^*) \right\}.
\end{align*}
Each one is solved by means of standard adaptive
routines~\footnote{See the routine \texttt{QAWFE} of the library QUADPACK at
\emph{http://www.netlib.org}.}; similar decompositions apply to
the other quantities of interest.\newline
\emph{Historical simulation}. A sample of $T_{MC}$ realizations of the risk factors
$X \sim N(0,\Sigma)$ is drawn and, correspondingly, the time series
of the portfolio variations $V_{t=1,\dots,T_{MC}}$ is computed. If $\tilde{V}_t$ are
the entries of $V_t$ sorted in ascending order, and assuming $t^* = T_{MC}\times \pstar$
to be integer, the historical VaR at the
significance level $\pstar$ is defined as~\footnote{If $T_{MC}\times \pstar$ is not
integer, $\Delta^*$ can be defined as the average $(-V_{s}-V_{u})/2$,
with $s,u$ being the two integers closest to $t^*$.}
\begin{equation}
  \Delta_{\h}^* = -\tilde{V}_{t^*}~.
  \label{eq:histo_var}
\end{equation}
The ES is obtained as the average on the left tail of the empirical
distribution
\begin{equation}
  E_{\h}^* = -\frac{1}{t^*} \sum_{t=1}^{t^*} \tilde{V}_t~.
  \label{eq:histo_ES}
\end{equation}
Confidence intervals for VaR are obtained from a basic result of order statistics. Let us
consider the sample of $T_{MC}$ i.i.d. deviates $V_i$ and indicate with $Q_p$
the $p$-th percentile of their distribution; then, the probability of $Q_p$ being
enclosed in $(\tilde{V}_{t^-},\tilde{V}_{t^+})$ is given by the following sum
of binomial probabilities~\cite{David:2003}
\begin{equation}
  P(\tilde{V}_{t^-} < Q_p < \tilde{V}_{t^+}) = \sum_{k=t^-}^{t^+-1}
  \binom{T_{MC}}{k} p^k (1-p)^{T_{MC}-k}~.
  \label{eq:bino_prob}
\end{equation}
To find the confidence interval associated to $\Delta^*_{\h}$ for a given Confidence Level (CL), we
find indices $t^{\mp}$ satisfying
\begin{equation*}
  P(\tilde{V}_{t^-+1} < Q_{\pstar} < \tilde{V}_{t^+}) \le CL
  \le P(\tilde{V}_{t^-} < Q_{\pstar} < \tilde{V}_{t^+})
\end{equation*}
where $Q_{\pstar}=-\Delta^*$ and the choice between possible different pairs $(t^-,t^+)$
satisfying the previous inequality is made requiring the confidence interval to
be as symmetric as possible around $\Delta^*_{\h}$. With these positions, Historical
VaR is estimated by
\begin{equation}
  (\Delta^*_{\h})^{+\delta^+}_{-\delta^-}~,\quad
  \left\{ \begin{aligned}
    \delta^+&\doteq -\tilde{V}_{t^-}-\Delta^*_{\h}\\
    \delta^-&\doteq \Delta^*_{\h} + \tilde{V}_{t^+}
  \end{aligned} \right.
  \label{eq:var_errors}
\end{equation}
with confidence level $CL$.

Since $E^*$ is a monotonously increasing function of $\Delta^*$, a lower and upper
bound for ES are easily obtained by evaluating the average in equation~\eqref{eq:histo_ES}
for $t^*=t^-$ and $t^*=t^+$ respectively. At the same CL as above, we estimate
\begin{equation}
  (E^*_{\h})^{+e^+}_{-e^-}~,\quad
  \left\{ \begin{aligned}
    e^+&\doteq -\left(\frac{1}{t^-}\sum_{k=1}^{t^-}\tilde{V}_k\right)-E^*_{\h}\\
    e^-&\doteq E^*_{\h} + \left(\frac{1}{t^+}\sum_{k=1}^{t^+}\tilde{V}_k\right)
  \end{aligned} \right. ~.
  \label{eq:es_errors}
\end{equation}

The sensitivities of $\Delta^*_{\h}$ are evaluated by
approximating its derivatives with respect to the remapped parameters
of equation~\eqref{indipDGN}, with the finite difference formula
\begin{equation}
  \left( \partial_{\beta} \Delta^* \right)_{\h} \doteq
  \left( \frac{\partial \Delta^*}{\partial \beta} \right)_{\h} =
  \frac{\Delta^*_{\h}(\beta+ \Delta\beta)
  -\Delta^*_{\h}(\beta- \Delta\beta)}{2 \Delta\beta}
  \label{eq:finite_diff}
\end{equation}
where $\Delta^*_{\h}(\beta\pm\Delta\beta)$ denotes the historical estimates for
the VaR obtained simulating the DGN portfolio after giving a positive/negative shock 
to the parameter $\beta$ while keeping fixed all the others. The error affecting
$\left( \partial_{\beta} \Delta^* \right)_{\h}$ is obtained by linear propagation
of the Monte Carlo errors on $\Delta^*_{\h}(\beta\pm\Delta\beta)$ through
equation~\eqref{eq:finite_diff}. An analogous procedure is applied to estimate
$\left( \partial_{\beta} E^* \right)_{\h}$ and its confidence interval.

The results discussed in the following sections have been obtained simulating
$T_{MC}=10^7$ synthetic portfolio scenarios for $N=15$ risk factors; for simplicity,
we simulated the remapped factors $Y\sim N(0,\mathbb{I})$ and the values
$V_t$ were constructed from equation~\eqref{indipDGN} after arbitrarily fixing
$\theta$, $\delta$ and $\lambda$. Attention has to be paid here to freeze the values of
$Y_t$, so that re-evaluation of the portfolio with shocked parameters is always carried
out based on the same sample. More details about efficient algorithms to
generate the scenarios under a quadratic approximation of the portfolio losses
can be found in~\cite{Glasserman:2002}.
The semi-analytical curves have been obtained by Fourier inversion of
equations~\eqref{VaRFFT}, \eqref{eq:ES}, \eqref{VaRderiv} and \eqref{ESderiv}, the
characteristic function and its derivatives as given by \eqref{charf} and~\eqref{derCF}. 
Given a grid of $\Delta^*$ values, the
corresponding $\pstar$, $E^*$, $\partial_{\beta}\Delta^*$ and $\partial_{\beta}E^*$ are
computed and compared with those from the historical simulation.
If one is interested in $\Delta^*$ corresponding to a precise spot value
of $\pstar$, inversion of equation~\eqref{VaRFFT} has to be nested in a root finding
procedure. Alternatively, the grid of $\Delta^*$ can be tuned in such a way
to obtain $\pstar$ values close enough to the required one.

\subsection{The lowest eigenvalue is negative}
In order to discuss the case $\lambda^*<0$, we fix the DGN parameters as follows:
$\theta=0$, $\delta_{i=1,\cdots,15}=1$, $\lambda^*=\lambda_{i=1,\cdots,5}=-2$,
$\lambda_{i=6,\cdots,9}=1$ and $\lambda_{i=10,\cdots,15}=2$ (CASE 1).

\begin{figure}
  \resizebox{1.\columnwidth}{!}{%
  \includegraphics{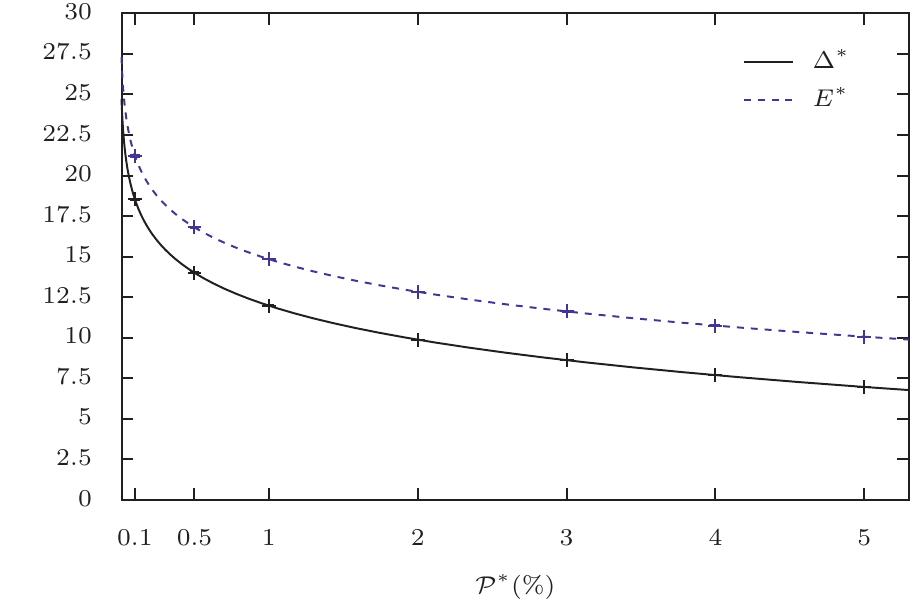}
  }
  %\vspace{5cm}       % Give the correct figure height in cm
  \caption{(Color online) CASE 1 ($\lambda^* < 0$): VaR and ES. The semi-analytical curves
  are compared to the points obtained via historical simulation for the values
  of $\pstar$ usually considered in practice.}
  \label{fig:x15mVaRES}
\end{figure}

\begin{figure}
  \resizebox{1.\columnwidth}{!}{%
  \includegraphics{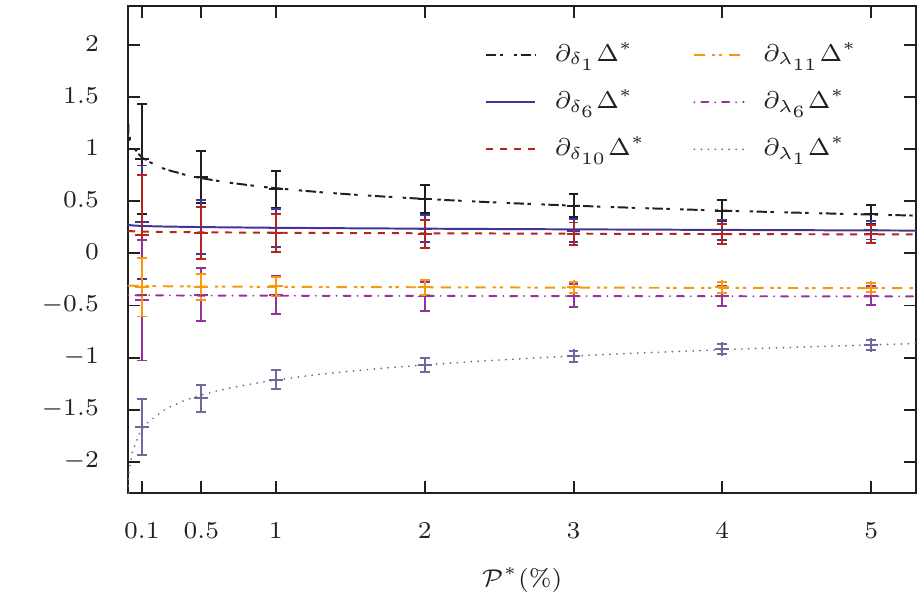}
  }
  \caption{(Color online) CASE 1: Value at Risk sensitivities.}
  \label{fig:x15mVaRder}
\end{figure}

\begin{figure}
  \resizebox{1.\columnwidth}{!}{%
  \includegraphics{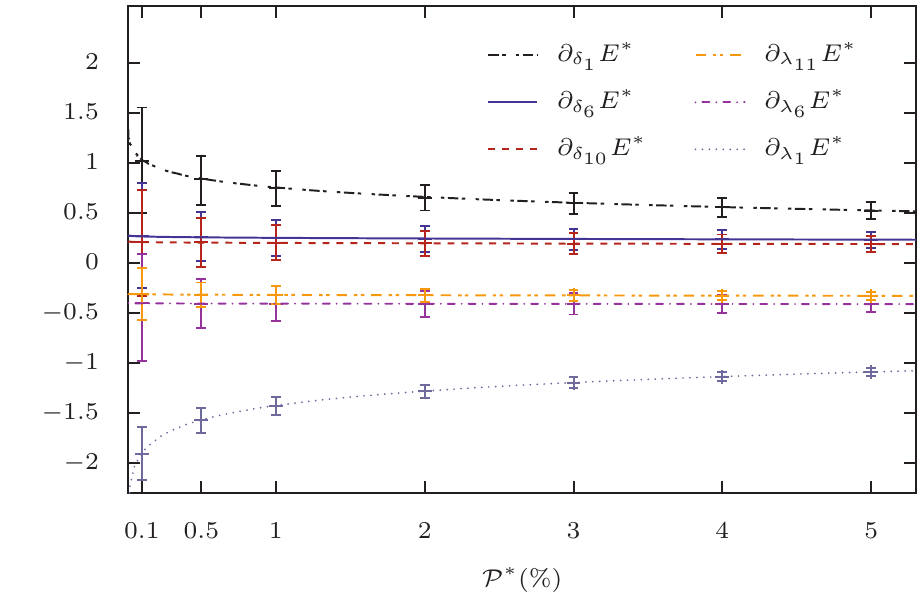}
  }
  \caption{(Color online) CASE 1: Expected Shortfall sensitivities.}
  \label{fig:x15mESder}
\end{figure}

The correspondence $\pstar$-VaR and $\pstar$-ES
as obtained from equations~\eqref{VaRFFT},\eqref{eq:ES} is shown in Figure~\ref{fig:x15mVaRES}.
As expected from their definitions, $E^*$ estimates are always larger than $\Delta^*$,
the shift vanishing in the limit of $\pstar$ approaching $0$.
The sensitivities for this set of parameters are illustrated in Figure~\ref{fig:x15mVaRder}
and Figure~\ref{fig:x15mESder}. With the exception of $\beta=\theta$, the
derivatives of $f(\phi)$ always depend on the pair $(\delta_i,\lambda_i)$; for
this reason the only cases we need to consider are $\beta=\theta,\delta_{1,6,10},\lambda_{1,6,10}$.
The sensitivities with respect to the central value $\theta$ are identically equal to $-1$; this result
can be verified directly from equations~\eqref{VaRderiv},\eqref{ESderiv} after substitution of
$\partial_{\theta}f(\phi)$ and reminding the definition of $\pstar$ and $p(\Delta^*)$, and
the corresponding curves are not shown for the sake of clarity.

All the curves are superimposed to the points obtained through historical simulation for
the interesting values of $\pstar=0.001,0.005,0.01,0.02,0.03,0.04,0.05$, along with
the corresponding error bars for $CL=98\%$; central points and error bounds
correspond to the definitions \eqref{eq:var_errors}, \eqref{eq:es_errors} for $\Delta^*$
and $E^*$ and to their extensions for the sensitivities.
The latter exhibit larger uncertainty, due
to the propagation of the statistical error, which is magnified by the finite difference
formula \eqref{eq:finite_diff} by a factor of order $\mathcal{O}(1/\Delta \beta)$. It
is also clear how this uncertainty increases for smaller $\pstar$, due to the decreasing size
of the sample on the left tail of the distribution of the generated values $V_t$.
On the whole, the full agreement with the Monte Carlo outcomes proves the effectiveness
of our analytical results and the reliability of their numerical implementation.

\begin{figure}
  \resizebox{1.\columnwidth}{!}{%
  \includegraphics{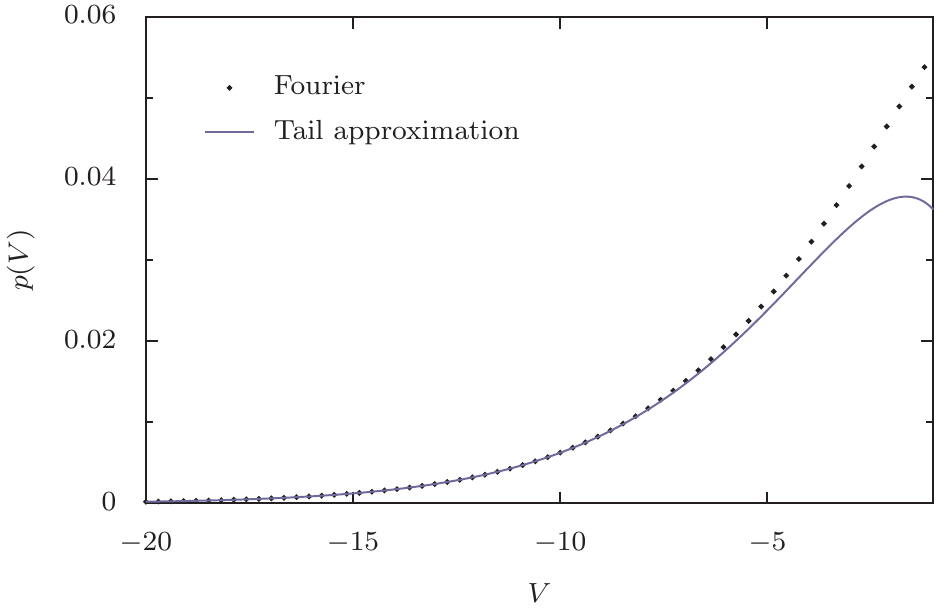}
  }
  \caption{(Color online) CASE 1: left tail of the PDF fitted by its analytical approximation corresponding
  to exponential decay.}
  \label{fig:x15mPDF}
\end{figure}

As discussed in section~\ref{sec:Risk_measures}, in this case the distribution function
$p(V)$ has both left and right exponential tails. This scaling is confirmed by
Figure~\ref{fig:x15mPDF}, where the left tail of the PDF, reconstructed by Fourier inversion
of equation~\eqref{GenFourT}, is fitted by its asymptotic approximation~\eqref{eq:Jaschke_m}.


\subsection{The lowest eigenvalue is zero}
\begin{figure}
  \resizebox{1.\columnwidth}{!}{%
  \includegraphics{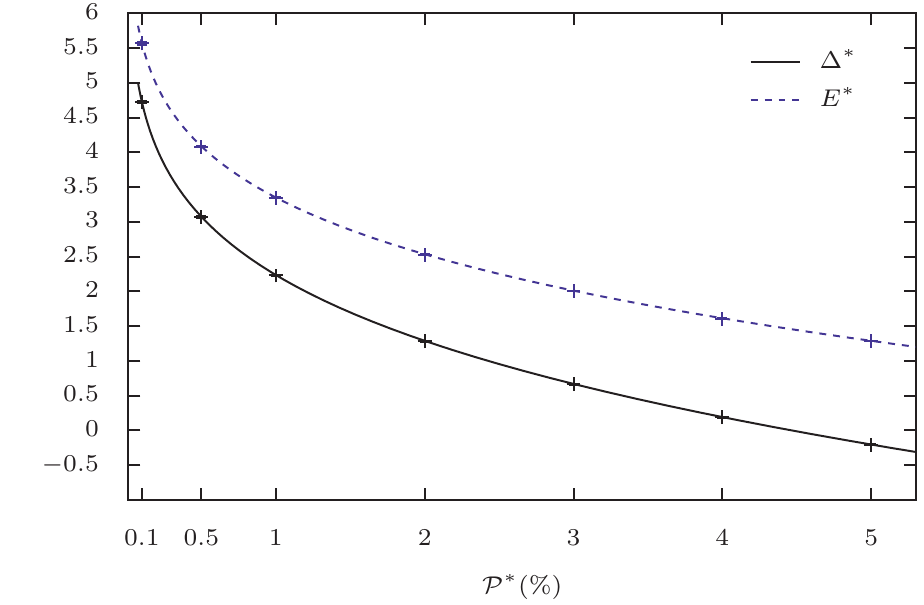}
  }
  \caption{(Color online) CASE 2 ($\lambda^* = 0$): Value at Risk and Expected Shortfall.}
  \label{fig:x15zVaRES}
\end{figure}

\begin{figure}
  \resizebox{1.\columnwidth}{!}{%
  \includegraphics{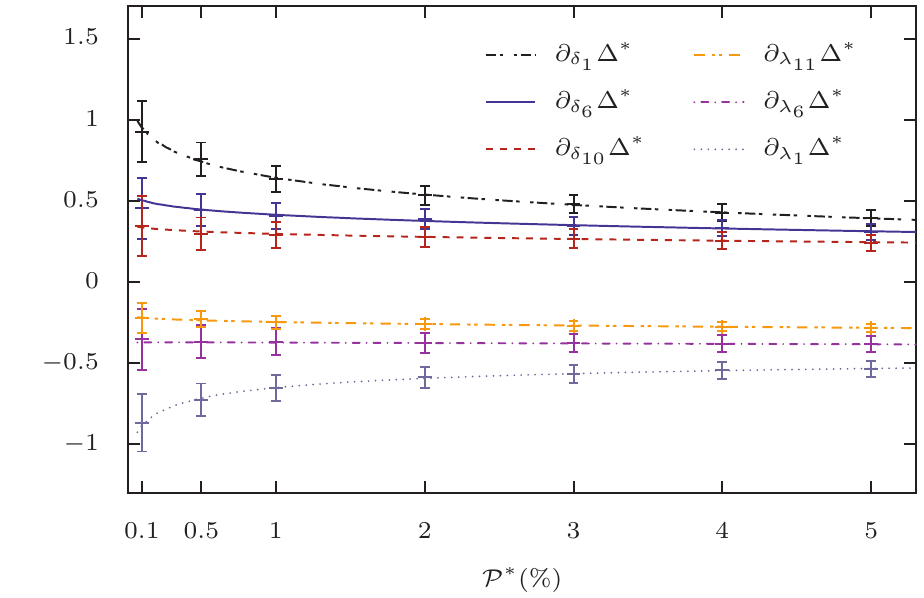}
  }
  \caption{(Color online) CASE 2: Value at Risk sensitivities.}
  \label{fig:x15zVaRder}
\end{figure}

\begin{figure}
  \resizebox{1.\columnwidth}{!}{%
  \includegraphics{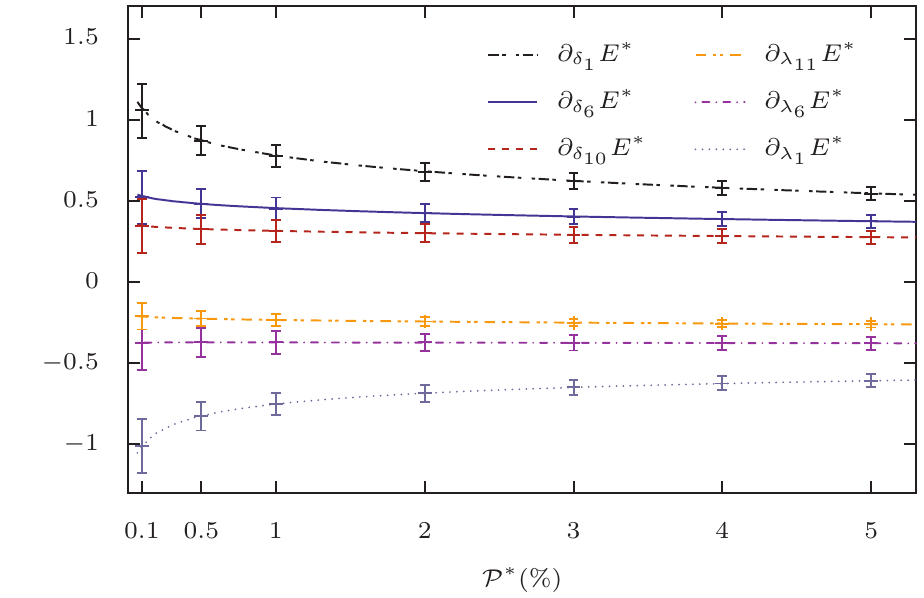}
  }
  \caption{(Color online) CASE 2: Expected Shortfall sensitivities.}
  \label{fig:x15zESder}
\end{figure}

\begin{figure}
  \resizebox{1.\columnwidth}{!}{%
  \includegraphics{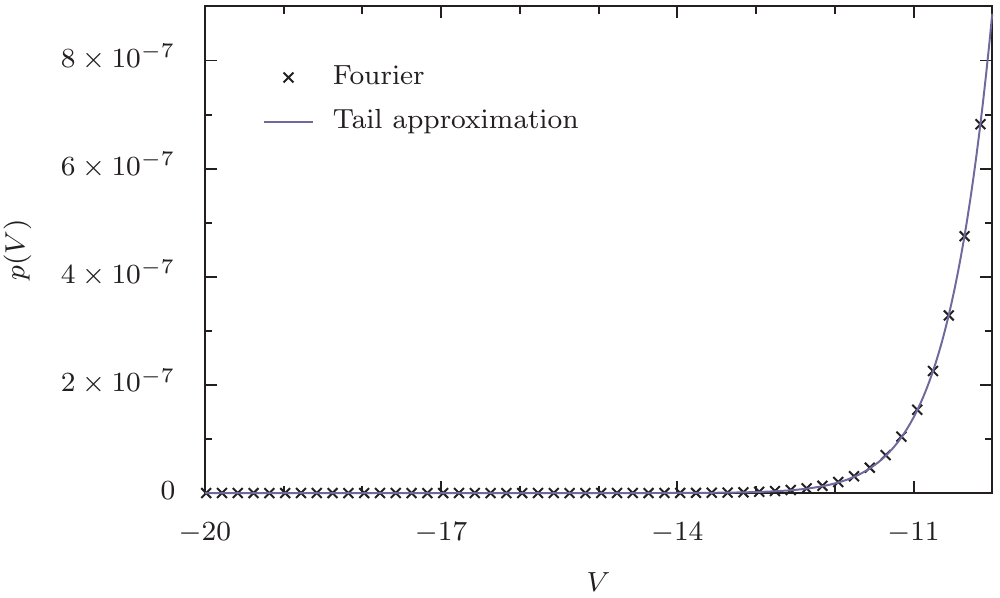}
  }
  \caption{(Color online) CASE 2: For $\lambda^* = 0$ the left tail of $p(V)$
  exhibits a Gaussian decay as expected from~\eqref{eq:Jaschke_z}.}
  \label{fig:x15zPDF}
\end{figure}

The case $\lambda^*=0$ is discussed here for the set of parameters:
$\theta=0$, $\delta_{i=1,\cdots,15}=1$, $\lambda^*=\lambda_{i=1,\cdots,5}=0$,
$\lambda_{i=6,\cdots,9}=1$ and $\lambda_{i=10,\cdots,15}=2$.
Figure~\ref{fig:x15zVaRES}-\ref{fig:x15zESder} show the same curves described
above, again exhibiting full statistical agreement with the Monte Carlo simulation.
Previous considerations about the statistical errors still apply and the only
significant difference is the shape of the PDF in Figure~\ref{fig:x15zPDF},
showing a Gaussian decay of the left tail, as expected from the
asymptotic expression~\eqref{eq:Jaschke_z}.
\subsection{The lowest eigenvalue is positive}
\begin{figure}
  \resizebox{1.\columnwidth}{!}{%
  \includegraphics{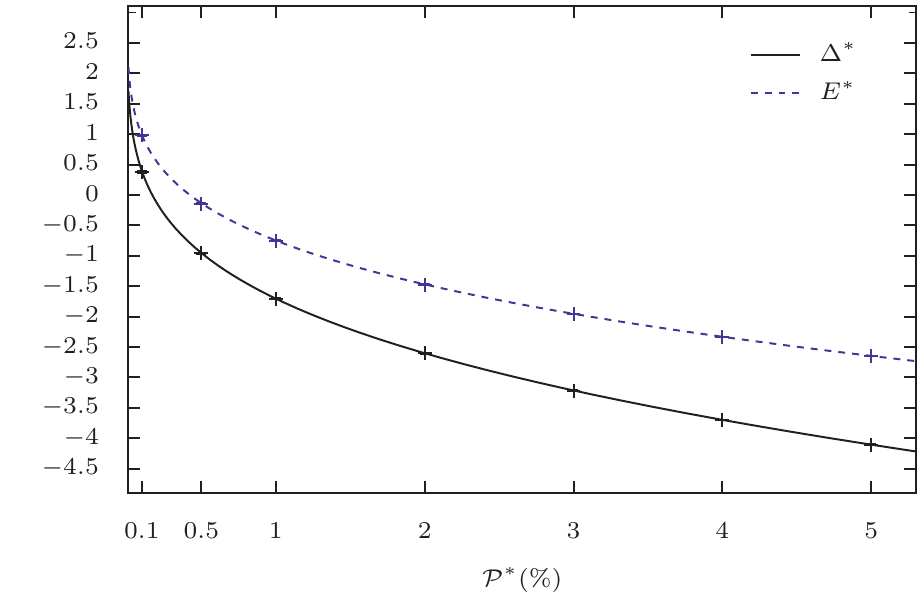}
  }
  \caption{(Color online) CASE 3 ($\lambda^* > 0$): Value at Risk and Expected Shortfall.
  The truncation of the PDF near the origin is responsible for the negative values for
  $\Delta^*$ and $E^*$ over a wide interval of $\pstar$, see the main text.}
  \label{fig:x15pVaRES}
\end{figure}

\begin{figure}
  \resizebox{1.\columnwidth}{!}{%
  \includegraphics{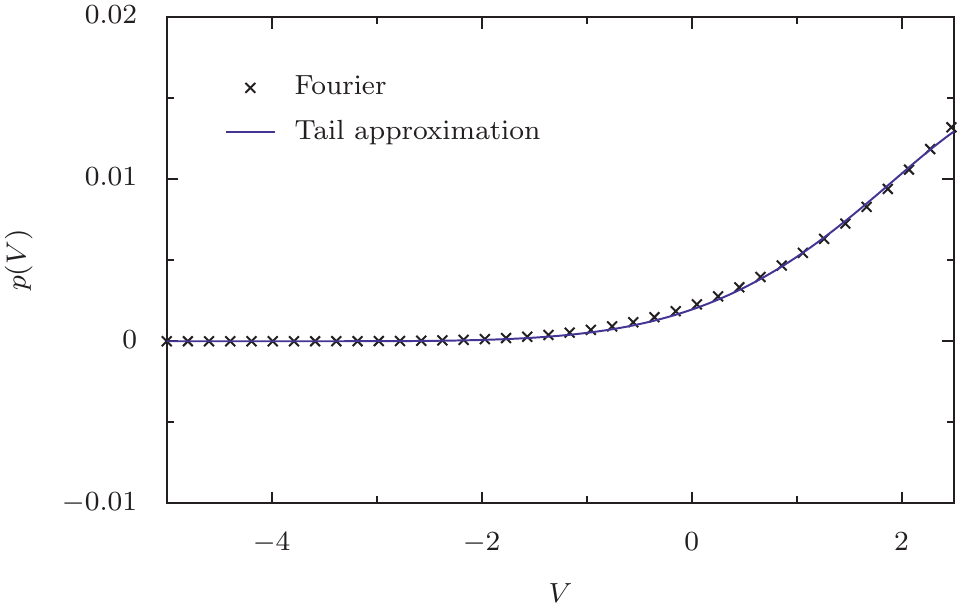}
  }
  \caption{(Color online) CASE 3: The PDF is limited from the left, decaying as
  a power law of exponent $N/2-1=6.5$.} 
  \label{fig:x15pPDF}
\end{figure}

The last case we consider corresponds to the choices: $\theta=0$, $\delta_{i=1,\cdots,15}=1$,
$\lambda^*=\lambda_{i=1,\cdots,4}=1$ and $\lambda_{i=5,\cdots,15}=2$ (CASE 3).
Since $\lambda^*>0$, the PDF of $V$ is limited from the left, see Figure~\ref{fig:x15pPDF};
its support is $[V_{\mathrm{inf}},+\infty)$ with $V_{\mathrm{inf}}=\theta-\sum_{k=1}^n
\frac{\bar{\delta^2_k}}{2\lambda_{i_k}}=-4.75$, and the left tail approaches $0$ with power
law scaling $(V-V_{\mathrm{inf}})^{N/2-1}$, see equation~\eqref{eq:Jaschke_p}.
As a consequence of the truncation near the origin of the $V$ axis, $\Delta^*$ and $E^*$
are negative for a quite large interval of $\pstar$ of interest, as illustrated in
Figure~\ref{fig:x15pVaRES}.

\begin{figure}
  \resizebox{1.\columnwidth}{!}{%
  \includegraphics{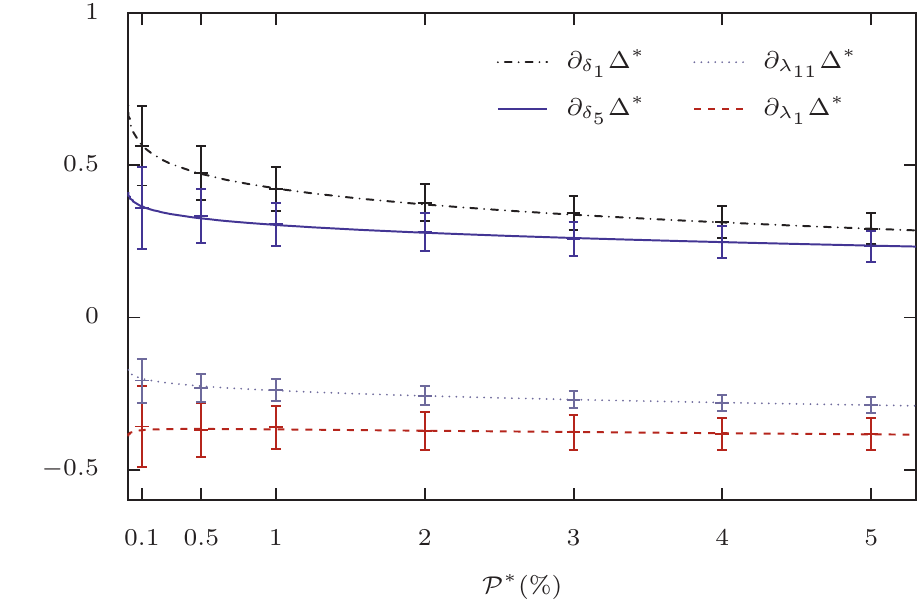}
  }
  \caption{(Color online) CASE 3: Value at Risk sensitivities.}
  \label{fig:x15pVaRder}
\end{figure}

\begin{figure}
  \resizebox{1.\columnwidth}{!}{%
  \includegraphics{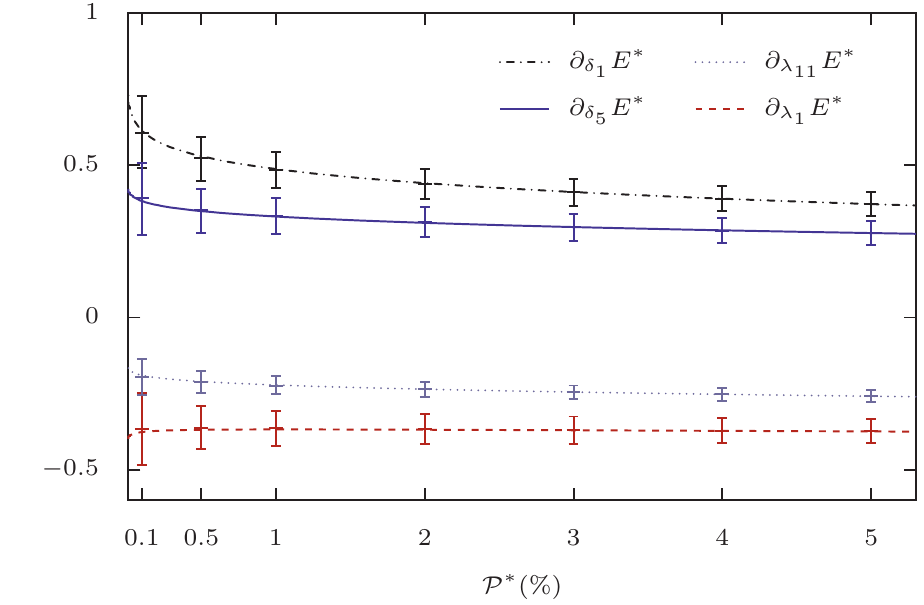}
  }
  \caption{(Color online) CASE 3: Expected Shortfall sensitivities.}
  \label{fig:x15pESder}
\end{figure}

For this set of parameters the unique $(\delta,\lambda)$ pairs to be considered are only
$(\delta_1,\lambda_1)$ and $(\delta_1,\lambda_5)$, corresponding to
the sensitivities reported in Figures~\ref{fig:x15pVaRder}, \ref{fig:x15pESder}.
%%%
\section{Conclusions}
\label{sec:conclusions}
In this paper, we addressed the problem of the evaluation of market risk exposure for
non linear portfolios, when price changes are approximated according to the Delta
Gamma Normal model. Exploiting the generalized Fourier representation of the probability
density function for the model, semi-closed form expressions for the Value at Risk and
the Expected Shortfall were derived. Similarly analogous expressions in terms of a
single Fourier integral were obtained for their first order derivatives;
thus our approach turns out to be especially useful, since these quantities are
of major practical relevance for the purposes of asset allocations and market risk
hedging, giving crucial information about the sensitivity of the portfolio with
respect to its risky components. It is worth mentioning that our approach could be
readily extended to compute higher order derivatives.
All of our formulae lend themselves to efficient numerical evaluation; they have been
tested by simulating synthetic portfolio scenarios via Monte Carlo.
This comparison, while confirming the reliability of our expressions with
a full statistical agreement, highlights their potential in practical applications,
since full Monte Carlo evaluation would require very large samples to obtain
accurate risk estimates. So, the availability of (semi) analytical techniques might
be welcome.

A different, yet related, problem stems from dealing with
time series of limited depth, as it is the case in financial practice. As far as the
historical approach is concerned, this implies a great statistical uncertainty
on empirical quantiles, while for our approach it would be reflected in a noisy 
estimation of the covariance matrix. We did not address this problem in the present
work, leaving it as a topic of possible future research. A further perspective would
be to extend our formalism to different financial contexts, where the characteristic
function may be known but explicit forms for the density function are not available,
which is especially true for many credit risk models.

\section*{Acknowledgements}

We would like to thank G. Montagna and O. Nicrosini for helpful discussions and
for reading the preliminary version of our manuscript.

\end{document}